# Sub-micron spin-based magnetic field imaging with an organic light emitting diode

Rugang Geng, Adrian Mena, William J. Pappas, Dane R. McCamey*

ARC Centre of Excellence in Exciton Science, School of Physics, UNSW Sydney, NSW 2052, Australia

*dane.mccamey@unsw.edu.au

**Abstract:**

Quantum sensing and imaging of magnetic fields has attracted broad interests due to its potential for high sensitivity and spatial resolution. Common systems used for quantum sensing require either optical excitation (e.g., nitrogen-vacancy centres in diamond, atomic vapor magnetometers), or cryogenic temperatures (e.g., SQUIDs, superconducting qubits), which pose challenges for chip-scale integration and commercial scalability. Here, we demonstrate an integrated organic light emitting diode (OLED) based quantum sensor for magnetic field imaging, which employs spatially resolved magnetic resonance to provide a robust mapping of magnetic fields. By considering the monolithic OLED as an array of individual virtual sensors, we achieve sub-micron magnetic field mapping with field sensitivity of ~160 µT Hz$^{-1/2}$ µm$^{-2}$. Our work demonstrates a chip-scale OLED-based laser free magnetic field sensor and an approach to magnetic field mapping built on a commercially relevant and manufacturable technology.

## Introduction

Magnetic field sensing and mapping are important for many scientific and technological applications across both physical[1-6] and biological systems[7-9]. Compared to classical methods, quantum sensing techniques have decisive advantages, including high sensitivity in field detection and high spatial resolution in field mapping. Among the many quantum techniques[3,10], nitrogen-vacancy (NV) centres in diamonds have emerged as an outstanding sensor platform, and achieved room-temperature picotesla level sensitivities[11], nanoscale spatial resolution[12-14] and good integration and miniaturization[15,16]. Meanwhile, organic semiconductors (OSCs) are proven to be extremely sensitive to magnetic fields[17-23], and OSC-based solid-state devices have been proposed as a new type of quantum sensor[22]. Unlike NV-based techniques, OSC-based magnetic sensors do not require optical pumping; moreover, they can provide both electrical and optical readout via optically and electrically detected magnetic resonance (ODMR and EDMR, respectively). EDMR allows for chip-scale integration of a single point-like sensor. Additionally, ODMR allows simultaneous acquisition of optical signals in the field of view, enabling spatially resolved sensing and imaging. OSC's are also inherently compatible with mass-produced consumer electronics, providing a potential pathway for ubiquitous deployment.

In this Article, we demonstrate an integrated solid-state device to detect and image magnetic field, where an organic light emitting diode (OLED) and a microwave resonator are laterally integrated on the same substrate (Fig. 1a). This new device architecture allows one to measure the magnetic field both electrically (via EDMR) and optically (via ODMR). The device can act not only as a point sensor by measuring its bulk EDMR or ODMR response, but also as a virtual array of sensors by spatially resolving ODMR. The latter offers a route for fast magnetic mapping without point-to-point scanning, which may have potential applications in quantum magnetic sensing and imaging[4].

**Principle of magnetic resonance in OLED.** The principle of optically and electrically detected magnetic resonance is based on the spin-dependent recombination and dissociation dynamics of charge carriers in the OLED[24]. When positive and negative charge carriers are injected from anode and cathode, they initially bind together electrostatically to form polaron pairs in the emissive layer. These polaron pairs can have either singlet or triplet dominant character depending on their spin configuration and a transition occurs between them via spin mixing. Polaron pairs can further recombine to form singlet or triplet excitons or dissociate back to free charge carriers with a rate dependent on their singlet-triplet symmetry. Under magnetic resonance ($f = \gamma B_0$, where $f$ is the microwave frequency, $\gamma$ is the gyromagnetic ratio and $B_0$ is the applied magnetic field), the spin orientations of charge-carriers in polaron pairs are manipulated which causes a change of the population ratio of singlet to triplet pairs. The population change is eventually transferred to the electroluminescence (EL) and the current through the recombination and dissociation process, respectively, leading to a change in both EL and current.

## Results

**Integrated device and EDMR characteristics.** The device structure consists of two main components: an omega-shape microwave resonator, and a micron-size heterostructure OLED located at the centre of the resonator (Fig. 1b). The microwave resonator is electrically isolated from the OLED using two insulating layers. See the details of the fabrication process in Method. Figure 1(a) displays the resonator-integrated substrate before the fabrication of the OLED, and the inset shows the completed device where the OLED was turned on at current I = 500 nA with bright and uniform EL. We first tested the EDMR characteristics of the device. Figure 1(c) shows a typical EDMR spectrum where the microwave frequency is fixed (710 MHz) and the external magnetic field $B_0$ is swept. The EDMR signal (the change in the device current) reaches its maximum at $B_0 \approx 25.3(3)$ mT, matching the expected resonant frequency. The EDMR spectrum can be fit using two Gaussian functions, corresponding to the two charge carrier species in a polaron pair[25]. See details of the EDMR measurement in Method. To measure the externally applied magnetic field $B_0$, the microwave frequency is swept, and the current change is monitored with lock-in detection. The change of the current is measured as a function of the microwave frequency and reaches its maximum at the resonant frequency; the external field can then be easily found given $B_0 = f/\gamma$. Figure 1(d) shows a frequency-swept EDMR spectrum where the external field $B_0$ is fixed at ~25.2 mT, and the spectrum peak occurs at $f \approx 708.5$ MHz matching the applied magnetic field. The gyromagnetic ratio in our device is $\gamma$ = 28.03 (±0.0024) GHz/T, which is experimentally obtained through a linear fit as shown in fig. 1(e). The value is slightly different from the free-electron gyromagnetic ratio (28.025 GHz/T) due to the weak spin-orbit coupling of the charge carrier spin states[26,27]. In addition, the magnetic field response of the sensor is shown to be linear over more than two orders of magnitude in the frequency domain (40 MHz to 6.0 GHz). We note that the upper limit of the resonant frequency here is purely limited by the microwave source, and much higher resonance frequency can be achieved with compatible microwave sources[22,28]. The minimum field that can be detected is limited by the intrinsic hyperfine field of the OSCs, although this can be overcome by applying an offset magnetic field e.g. with a microwire[22,29,30]. In addition, for ultrasmall field detection, the influence of the geomagnetic field (~50 μT) on the spin charge carriers in OSCs needs to be considered[31], which can be potentially removed by magnetic shielding.

**Point sensor via EDMR.** To demonstrate the sensing capacity of the device, a permanent magnet is used to create a known imaging phantom. More details are discussed in Supplementary Method 4. Figure 2(a) sketches the experimental scheme, where the magnet is located next to the device and the OLED employed as a magnetic field point sensor via a frequency-swept EDMR measurement. We note that in OSCs the resonant peak frequency of the EDMR spectrum is only dependent on the strength of the external field B rather than its direction[18], and this directional independence is also observed in our device (More details in Supplementary Method 3). Here we carry out two independent measurements where the device is stepped along the x- (horizontal) and y-direction (vertical). At each step, the microwave frequency is swept, and the magnitude of B determined from the resonant frequency of the resulting EDMR spectrum. The measurement results show that the sensor can detect magnetic field over a broad range with high accuracy, where the similarity between the experimental result and a computational simulation of the expected field is about 99.8 % (Fig. 2b) and 98.6 % (Fig. 2c), respectively. See more details in Supplementary Method 4.

**Spatially resolved ODMR and magnetic field mapping.** Our integrated device is not only capable of sensing magnetic fields electrically via EDMR, but also provides optical accessibility for magnetic field mapping via a spatially resolved ODMR. To measure the spatially resolved ODMR, an optical microscope is used to image the device onto an sCMOS camera (Fig. 3a). A square-wave microwave signal (0.5 Hz) is applied, and the difference in EL between the on and off cycles is measured with the camera. The microwave frequency is swept, and a similar image taken at each frequency. Each pixel of the camera therefore measures an ODMR spectrum associated with a spatial region of the OLED. See more details in the Methods. To improve the signal-to-noise ratio (SNR) of the ODMR spectrum so that the magnetic field can be more precisely measured, the camera pixels are binned to form super-pixels (see Fig. 3b). Figure 3(c) shows the ODMR spectrums of two individual super-pixels at separate locations. Though the SNR of the data is still relatively low (~4) after binning, the spectrum can be well fit using a double Gaussian function, from where the magnetic field is acquired by converting the resonant frequency ($f_{ODMR}$) to a field strength ($|B|$). Figure 3(d) shows a 2D map of the measured magnetic field across the entire region (152.5 μm × 152.5 μm) with super-pixel size of ~ 0.91μm (binning size n =3). We note that the super-pixel size here is above the optical diffraction limit of the microscope objective ($\lambda/(2NA) = 714\ nm$) for a typical EL wavelength of λ = 600 nm.

The measured field shows a clear and smooth gradient change along the x-direction while remaining the same along the y-direction, which is consistent with the orientation of the magnet. We note that there is a barely visible ring feature in the centre of the field map, which is due to the via in the dielectric which defines the OLED active area. The EL signal and the related ODMR spectrum is observed across the whole field of view, which is much larger than the OLED region (D=80 μm) defined through the photolithography process (see Supplementary Fig. 1f). The reason is due to the high hole conductivity in the PEDOT:PSS (>1,000 S cm$^{-1}$)[32-34], as a result, the injected holes from the ITO electrode diffuse in the PEDOT:PSS layer along the in-plane direction, leading to a EL emission over a much larger area. The SNR of the ODMR spectrums differ inside and outside of this region (see Supplementary Fig. 8). Figure 3(e) shows a zoom-in view of a local region (9.1 μm × 9.1 μm) of the 2D map in figure 3(d) with a variety of binning size. As the binning size increases, the spatial resolution of the field mapping decreases; accordingly, the standard error of the fit decreases, indicating the enhancement of the measurement sensitivity. See more details in Supplementary Figure 8.

We now turn to the relationship between the magnetic field sensitivity and the spatial resolution of the field mapping. In general, the magnetic field sensitivity is defined as the minimum detectable field $B_{min}$, which corresponds to the error of an individual measurement. By combining the signal from neighbouring pixels, the measurement error can be reduced by $\sqrt{N} = n$ times, where N is the total number of camera pixels in each individual super-pixel with binning size $n$ ($N = n \times n$), although this also reduces the spatial resolution. To account for the measurement time, the sensitivity is represented as $\eta = B_{min} \times \sqrt{T} = SE(n) \times \sqrt{T}$, where $SE(n)$ is the standard error ($SE$) of a single ODMR measurement of an individual super-pixel with binning size $n$, and T is the total data acquisition time for each microwave frequency step. We note that T = 400 s is used in the actual experiment (see Supplementary Method 6), and the $SE$ of the measurement can be extracted from the fit of the ODMR spectrum. For magnetic field mapping with spatial resolution of ~0.91(5) μm (binning size n = 3) in Fig. 3(d), the magnetic field sensitivity is ~ 233.04 μT Hz$^{-1/2}$ in the OLED region and ~ 163.16 μT Hz$^{-1/2}$ in the diffusion region. When the super-pixel size is increased to ~14.64 μm (binning size n = 48), the sensitivity is improved to ~ 136.88 μT Hz$^{-1/2}$ in the OLED region and ~ 40.75 μT Hz$^{-1/2}$ in the diffusion region. Although the measured sensitivity follows the general $1/n$ rule (see Supplementary Fig. 11), the actual improvement ratio of the sensitivity (e.g., 163.16 ÷ 40.75 ≈ 4 in diffusion region) is much smaller than the trade-off ratio of the spatial resolution 48 ÷ 3 = 16 times). We suspect that this discrepancy arises due to device-related noises including the fluctuation of EL intensity caused by the electrical coupling between the OLED and the resonator (see Supplementary Figure 4), and other technical contributions, which suppress the SNR improvement which can be obtained via binning. The continuous wave (CW) ODMR shot-noise-limited sensitivity of our current setup is estimated to be 54.8 μT Hz$^{-1/2}$ μm$^{-2}$, which is about 3 times better than the measured sensitivity in this work. Details of the calculation can be found in the Methods.

Alongside magnetic field mapping, the device can also be used to measure magnetic field gradients at μm scales. The field gradient is defined as $G = \Delta B/\Delta x$, where $\Delta B$ is the difference of the measured magnetic field between two super-pixels with size of $w \times w$ acting as two virtual point sensors, and $\Delta x$ is the center-to-center distance between them. The averaged gradient along the x-direction in Fig. 3(d) is estimated as ~ 3.7 μT/ μm ($\Delta B \approx$ 555.7 μT, $\Delta x \approx$ 151.0 μm), while no clear gradient is observed inside the active region of the device along the y-direction due to the device alignment relative to the orientation of the magnet (see Fig. 2a). The minimum detectable field difference $(\Delta B)_{min}$ is determined by the magnetic field sensitivity (the minimum detectable magnetic field) which is dramatically impacted by the binning size; therefore, the field gradient sensitivity (or minimum detectable field gradient) is limited by the size of the virtual point sensors. In addition, the size of the virtual point sensors also sets a limit on the spatial resolution ($(\Delta x)_{min} = w$) of the field gradient. Based on error propagation, the magnetic field gradient sensitivity $\eta_G$ can be calculated as a function of virtual pixel size $w$ and the gap distance $\Delta x$. For the details of the calculation refer to the Supplementary Method 6. As shown in Fig. 4, the field gradient can be achieved at μm scale with a relatively good gradient sensitivity; the gradient sensitivity can be improved by either increasing the virtual pixel size or the gap distance or increasing both at the cost of spatial resolution. We note that all the noise sources that limit the magnetic field sensitivity discussed above will limit the gradient sensitivity as well.

**Discussion**

One significant challenge of using any resonance-based technique for magnetic field sensing is the measurement time to find the resonant frequency across a broad range, which can be time-consuming, particularly with small frequency step size and long averaging time for better SNR. We anticipate that operationally a coarse scan across a broad range followed by a fine scan across the resonance range can be used to shorten the measurement time. However, such an improvement may be limited by the resonance linewidth which sets the upper limit of the frequency step size. Another challenge is the field sensitivity in both EDMR and ODMR, especially the spatially resolved ODMR. This may be improved in a number of ways: firstly, by minimizing the sample-related noises, especially the electrical coupling between the resonator and the OLED through device architecture optimization. Secondly, the sensitivity can be further enhanced by employing coherent modulation techniques (e.g.

Ramsey or dynamical decoupling schemes)[14,35-38]. One of the motivating factors for this study was the presence of reasonably long spin phase coherence times of polarons in organic devices at room temperature – $T_2$ approaching 1 µs at room temperature has been observed in EDMR measurements, and the use of perdeuterated organic materials may be used to increase these times[18,39]. Concerted effort aimed at identifying or developing materials with even longer phase coherence times would seem to hold promise.

In contrast to NV-based detection where multiple resonant peaks may occur because of the different crystallographic axes of NV in diamond, there is only one resonant peak in our device sensor, sensitive to the strength of the external magnetic field regardless of the field orientation. This means that no alignment of the sensor is required in the detection of the field strength. This alignment-free characteristic can potentially be useful in applications where the magnitude rather than the direction of the field is of importance, and inaccuracy caused by improper orientation of the magnetometer can be reduced (i.e., Hall-effect probe). We note that the amplitude of the resonant signal is proportional to the projection of the microwave field $B_1$ along the static external field $B_0$, and that it reaches the minimum or even vanishes when $B_0$ is in parallel with $B_1$. More details are discussed in Supplementary Method 3. The sensitivity issues this causes can be potentially solved by generating a $B_1$ field with large directional inhomogeneity across the OLED so that there is always a portion of $B_1$ field that is projected orthogonal to $B_0$, leading to measurable resonances regardless of the orientation of $B_0$.

In order to detect the direction of the magnetic field, we can potentially extend the device architecture with two mutually perpendicular metallic strip lines integrated underneath the OLED[22]. This will provide an in-plane microwave field with arbitrary direction. Along with the out-of-plane microwave field from the already integrated resonator, there will be three independent microwave fields that are perpendicular to each other. By repeating the measurement with each microwave field, the corresponding vector components of the unknown magnetic field can be detected. We note that the use of the metallic strip lines may block the light emission and consequently limit the optical readout in ODMR. Additionally, these metallic layers including the top Al electrode of the device can distort the external magnetic field[28], posing more challenge for the precise vector measurement, especially at high magnetic fields.

## Conclusions

We have shown an OLED-based integrated quantum magnetometer capable of both optical and electrical readout. We emphasize that our device can work not only as a point sensor under electrical operation via bulk EDMR or ODMR, but also as a virtual array of sensors under optical readout via spatially resolved ODMR. Magnetic field mapping and a field sensitivity of 163.2 µT Hz$^{-1/2}$ with a spatial resolution of 0.91 µm is achieved in our current setup. This work provides a proof-of-concept demonstration of OLED-based magnetic field mapping. The field sensitivity can be further improved by several approaches, including minimizing device-related noise by optimizing the device architecture; improve the SNR of spatially resolved ODMR by utilizing spin coherent manipulation; and reducing the resonance linewidth using deuteration materials. Our work provides a chip-scale and laser-free OLED-based platform for magnetic field sensing and mapping, with potential applications for ubiquitous quantum sensing and mapping which leverages the significant investment in consumer OLED technologies.

## Acknowledgment

This work is supported by the Australian Research Council via the ARC Centre of Excellence in Exciton Science (CE170100026), and the Linkage Infrastructure, Equipment and Facilities scheme (LE150100075). A. M. and W. P. acknowledge the financial support from Sydney Quantum Academy. This work was performed in part using facilities of the NSW Node of the Australian National Fabrication Facility. We also acknowledge the facilities and the scientific and technical assistance of Microscopy Australia at the Electron Microscope Unit (EMU) within the Mark Wainwright Analytical Centre (MWAC) at UNSW Sydney.

**Author contributions**

D.R.M conceived the study. Devices were designed by R.G. and D.R.M. and fabricated by R.G. Experiments were undertaken by R.G., A.M. and W.J.P. All authors contributed to the analysis and preparation of the manuscript.

**Competing interests**

The authors, through the University of New South Wales, are pursuing a patent application based on the work described in this manuscript. All authors are inventors, and no other inventors are named. The application is currently being drafted for submission.

**Data Availability Statement**

The datasets generated during and/or analysed during the current study are available from the corresponding author on reasonable request.

## Methods

**Fabrication of the integrated microwave resonator.** To have the optical access to the device, the resonator structure and the OLED should be laterally separated so that the light can emit out from the ITO/glass side of the substrate. And the main challenge of integrating the resonator with an OLED on the same ITO-based glass substrate is how to electrically isolate them from each other. Here we employ low-temperature atomic-layer-deposition (ALD) method for the insulating layer depositions, providing conformal and high quality electrically insulating layer with thin thickness. The main procedures are as follows: **1)** prepatterned ITO (120 nm) on glass substrates (30.0 mm × 20.0 mm × 0.7 mm) was purchased from a commercial company. **2)** prepare the first insulating layer $Al_2O_3$ between the ITO layer and the following resonator layer. The geometry of the insulating layer was patterned through the standard photolithography process (MA6 system with negative photoresist nLOF2020 and developer ZA826MIF), and the $Al_2O_3$ (45 nm) layer was deposited by low-temperature ALD, followed by the lift-off process in NMP bath. **3)** prepare the resonator layer on top of the first insulating layer. The structure of the resonator was defined through the standard photolithography (the same as in step 2), and then metal layer of Ti (10 nm) / Au (500 nm) / Ti (10 nm) was thermally deposited in a thermal deposition chamber (Jurt J. Lesker) followed by standard lift-off process. The 10 nm Ti layers were adhesion layers. **4)** prepare the second insulating layer on top of the resonator the same way as for the first insulating layer in step 2. This second insulating $Al_2O_3$ (45 nm) layer is to electrically isolate the resonator itself from the top electrode of the OLED which was deposited in the later device fabrication process. The final layer structure of the resonator-integrated substrate is: bottom electrode layer of ITO (120 nm)/first insulating layer of $Al_2O_3$ (45 nm) / microwave resonator layer of Ti (10 nm) / Au (500 nm) / Ti (10nm) / second insulating layer of $Al_2O_3$ (45 nm). More details of the fabrication process are discussed in the Supplementary Method 1.

**Fabrication of micron-size OLED onto the resonator-integrated substrate.** The resonator-integrated substrate was firstly cleaned by using UV ozone cleaner (purchased from Ossila) for 10 minutes, followed by spin coating of PEDOT: PSS ((purchased from Heraeus, Al 4083) at 3000 rpm for 1 min, which was baked for 2 hours at 120 °C on hotplate, resulting in a film thickness of about 35 nm. The sample was then transferred to a glove box ($O_2$ < 0.5 ppm, $H_2O$ < 0.5 ppm) where the SY-PPV solution (3 mg/ml in toluene) was spin coated at 1200 rpm for 1 minute and then baked for 2 hours at 60 °C on hotplate, resulting in a film thickness of about 80 nm. The SY-PPV solution was filtered using a PTFE syringe filters with pore size of 0.45 μm before spin coating to remove the polymer aggregates. The extra part of the SY-PPV layer on the top of the Au resonator and the electrode pads was carefully removed by using cotton rod. Then sample was transferred to a high vacuum chamber (< $10^{-8}$ mbar) for the deposition of LiF (1 nm)/Al (100 nm) using a shadow mask. The shadow mask was carefully aligned with the substrate so that Al was deposited to the target area only (top of the second insulating layer region) to avoid any possible short-circuit connection between the resonator and the top Al electrode. After fabrication, the device was encapsulated with a thin glass lid with recessed cavity using UV-activated epoxy inside the glovebox. A thin desiccant sheet (as moisture and oxygen absorber) was sticked onto the inner surface of the recessed cavity to prevent the device degradation in the air.

**EDMR measurement set-up.** For the EDMR measurements in Fig. 1 and Fig. 2, the OLED was operated under a constant current of 0.5 μA (Keysight, SMU B2901A) at room temperature. The device was mounted onto a PCB via a 3D printed lid-base board, and the device was electrically connected to the PCB through pogo pins integrated on PCB, and the PCB was connected to all the measurement instruments using SMA cables. A photograph of the setup in details was demonstrated in SI. A signal generator (SRS SG396) was connected to the microwave-resonator, providing an input microwave signal which was pulse modulated with 10 μs pulse width and 10 kHz modulation. The other end of the resonator was connected to a 50 Ω terminator. During the EDMR measurement, the resulting periodic changes in the device current were first amplified through a low-noise current amplifier (SRS SR570) utilising a 6 dB bandpass filter at 10 kHz, and then were detected by the lock-in amplifier (SRS SR865A).

**Spatially resolved ODMR measurement.** As shown in Fig. 3(a), the device was mounted on a 3-axis optical stage and then well aligned with an optical imaging system. The light emitting out of the device from the ITO side, is collected by an Infinity Corrected objective (20X Mitutoyo Plan-Apochromat Objective, NA = 0.42, working distance = 20.0 mm, focus length = 10.00 mm), and then refocused onto a scientific CMOS camera (Andor iStar sCMOS 18U-A3 with working temperature of 0.0 °C) through a compatible tube lens (focal length =200.0 mm), where the EL intensity signal was detected and acquired by the camera. Under 20X magnification, the pitch between two adjacent pixels on the OLED plane is about 0.30(5) µm. The OLED was operated under a constant current of 0.5 µA (Keysight, SMU B2901A) at room temperature, and the resonator was connected to the signal generator (SRS SG396). A test magnet was located next to the device, providing a static external magnetic field for Zeeman energy splitting. The microwave field output was modulated by a 0.5 Hz square-wave sequence with 200 operation sequences, the EL intensity signal of the on and off cycle was recorded by the camera (exposure time of 980 ms) at each microwave frequency. The microwave frequency was swept, and eventually a full set of EL intensity data was recorded as a function of the microwave frequency. By calculating the averaged change of the EL signal between on and off cycle as a function of microwave frequency, we were able to obtain the ODMR spectrum at each camera pixel, namely a spatially resolved 2D ODMR spectrum.

**Shot-noise-limited sensitivity.** CW ODMR shot-noise-limited sensitivity is calculated in the following equation[35]: $\eta_{CW-ODMR} = \frac{8\pi}{3\sqrt{3}} \frac{\hbar}{g_e \mu_B} \frac{\Delta v}{C\sqrt{R}}$, where C is the contrast of the ODMR spectrum, R is the effective rate of photon detection of the OLED emission per µm$^2$, and $\Delta v$ is the linewidth of the ODMR resonance. In our measurement, the contrast C varies between 0.23 % to 0.55 % depending on the location. The camera pixel well depth P is $3\times10^4$ electrons, the exposure time $\tau_{exp} \sim$ 1.0 s for the maximum raw EL signal, and the peak quantum efficiency (QE) of the camera is about 50 %. Given the individual pixel size A (0.31 µm ×0.31 µm), the effective photon detection rate R per µm$^2$ is calculated as: R = P/( QE×A×$\tau_{exp}$) ≈ 6.24 ×10$^5$ photons/s per µm$^2$. For the ODMR linewidth (see Fig.1d), we select the narrower one $\Delta v = \sigma_1 \sim$ 6.15 MHz. The averaged contrast C is ~0.39 %. Therefore, by using these experimental values of C = 0.39 %, $\Delta v$ = 6.15 MHz and R = 6.24 ×10$^5$ /s (per µm$^2$), the shot-noise-limited sensitivity is calculated as $\eta_{CW-ODMR}$ = 54.80 µT Hz$^{-1/2}$ (per µm$^2$).

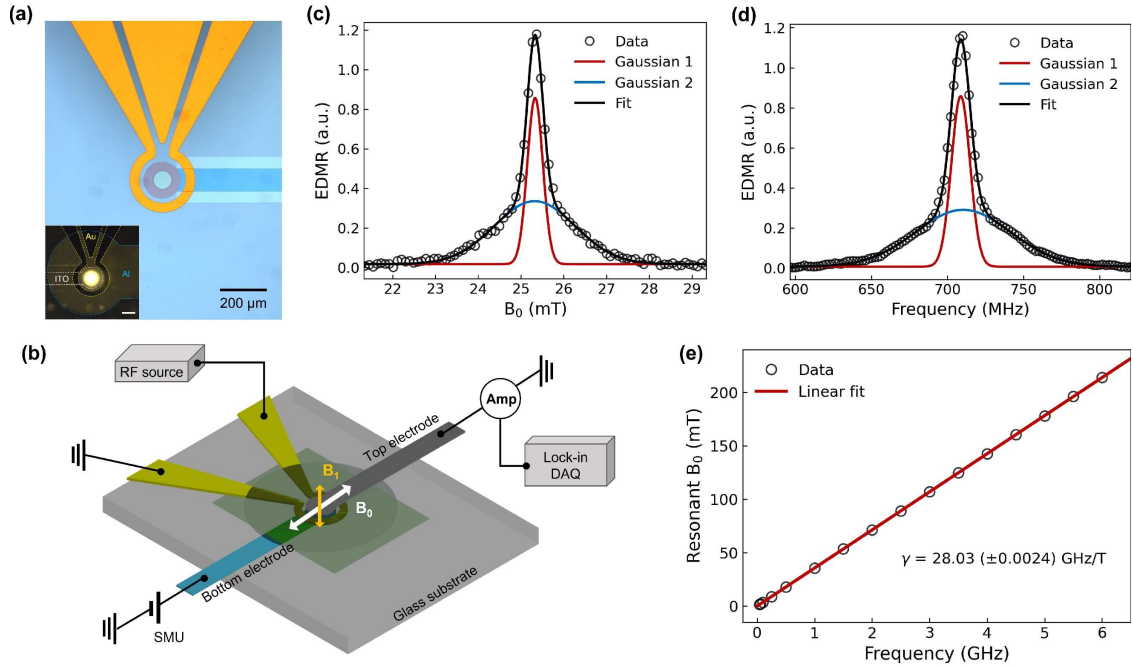

**Figure 1: Device structure, experimental setup, and EDMR characterization**. **(a)** Photograph of the integrated microwave resonator where an omega-shape resonator is integrated on the prepatterned ITO/glass substrate. The active area in the middle has a diameter of 80 μm, which is defined through photolithography and insulating layer deposition. The inset shows the photograph of an integrated OLED at current of $I$ = 500 nA (corresponding current density of ~10 mA/cm$^2$). **(b)** Sketch of the integrated device structure and the experimental measurement configuration, employed with the AC magnetic field $B_1$ created by the microwave resonator and the static magnetic field $B_0$ generated by an external electromagnet. **(c)** A conventional EDMR spectrum where the static magnetic field $B_0$ is swept with a fixed microwave frequency of 710 MHz. The spectrum is well described by the sum (black) of two Gaussian functions (red, blue), corresponding to the two hyperfine-field distributions ($\sigma_1$ = 0.18(2), $\sigma_2$ = 0.94(2)) experienced by the electron and hole spins, respectively. $\sigma_1$ and $\sigma_2$ represent the standard deviation of the two Gaussian functions. **(d)** A frequency-swept EDMR spectrum where the microwave frequency is swept with a fixed magnetic field $B_0 \approx$ 25.2(5) mT via fixing the current in the electromagnet. The spectrum can be well fitted using two Gaussian functions with standard deviation of $\sigma_1$ = 6.15(1) and $\sigma_2$ = 31.23(0), respectively. We note that the background noise caused by the frequency sweep is removed from the plots in (d). More details are discussed in Supplementary Method 2. **(e)** Plot of the maximum-peak value of the magnetic field $B_0$ in the EDMR spectrum as a function of the applied microwave frequency. A linear fit (red line) of the data yields a gyromagnetic ratio γ = 28.03 (±0.0024) GHz/T and a corresponding g-factor g = 2.0026 (± 0.00017).

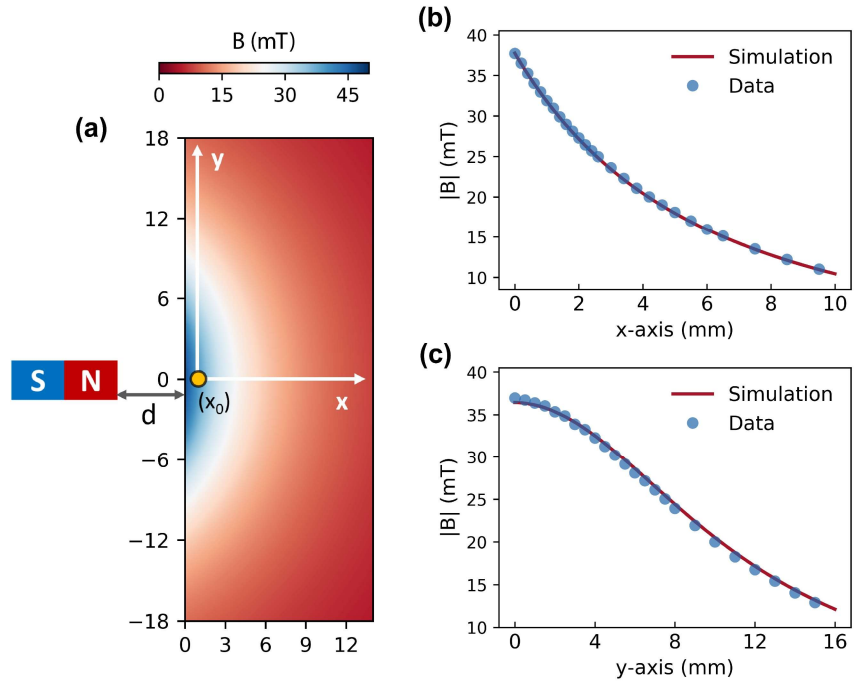

**Figure 2: EDMR-based magnetic field sensing. (a)** Sketch of the experimental setup (not to scale). A cylindrical magnet is located next to the device with the cylindrical axis of the resulting magnetic field aligned in the plane of the device substrate. 2D simulation of the spatial distribution of the decaying magnetic field strength generated by the cylindrical magnet in a region of 14.0 mm × 36.0 mm in the x-y plane with a distance of d = 10.0 mm from the magnet. The distance d corresponds to the half size of the device substrate width as the OLED is located at the center of the rectangular glass substrate (see Supplementary Figure 6). In actual experiments, we initially set a tiny gap ($x_0$) between the substrate edge and the magnet at the starting position to avoid possible physical contact between them during the movement. The total distance between the OLED (yellow dot) and the magnet is d + $x_0$. The x and y coordinates represent the horizontal and vertical movement directions in the laboratory frame, respectively. The OLED here works as a point detector to measure the magnetic field strength generated by the magnet, and $x_0$ represents the starting position of the measurement. **(b)** Magnetic field detection as the device is stepped along the x-direction. The magnetic field strength is measured via the frequency-swept EDMR spectrum at each position, and the solid curve is the simulation with an estimated starting position of $x_0 \sim 0.20$ mm. **(c)** Magnetic field detection as the device is stepped along the y-direction with an estimated starting position of $x_0 \sim 0.40$ mm.

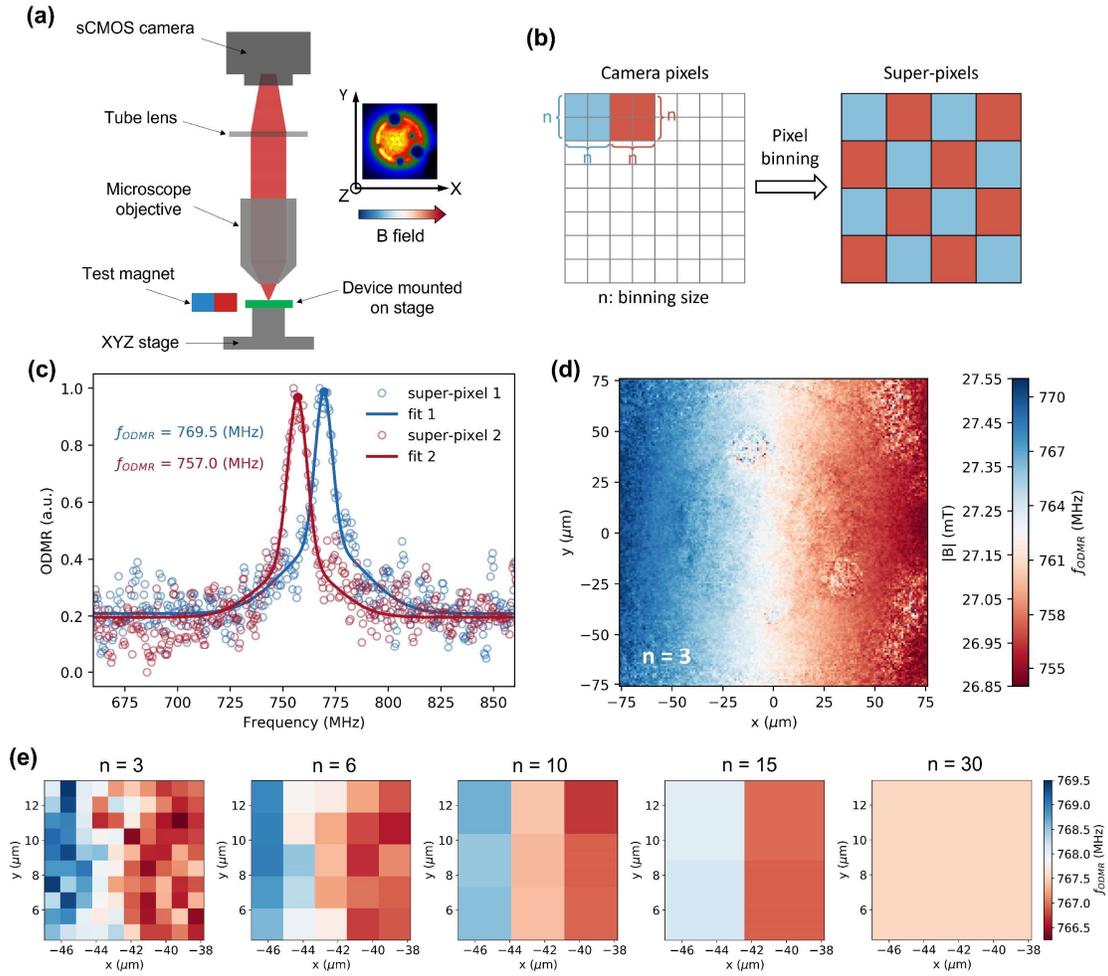

**Figure 3: Spatially resolved ODMR-based magnetic field mapping. (a)** Sketch of the setup for spatially resolved ODMR. The inset shows the image of EL intensity captured by the sCMOS camera. The B field arrow represents the magnetic field gradient across the OLED along x-direction in the horizontal x-y plane. **(b)** Scheme of pixel binning where n × n adjacent camera pixels are merged into one combined pixel called "super-pixel" via pixel binning process. The optical signal (EL intensity) of each super-pixel is the average of the signals of all the n × n individual camera pixels. **(c)** Double Gaussian fits of ODMR spectrums of two super-pixels with binning size n = 3. Super-pixel 1 and super-pixel 2 corresponds to the super-pixel at position of (-63.4 µm, 0.0 µm) and (52.4 µm, 0.0 µm) in (d), respectively. The solid circle dots label out the resonant peak position in the fit curves. **(d)** 2D spatial map of the resonance frequency ($f_{ODMR}$) of the ODMR spectrum of 166 × 166 super-pixels with binning size n = 3. The entire region contains 500×500 pixels, and the super-pixel size is about 0.91(5) µm × 0.91(5) µm (n = 3). Weak EL signal is also observed outside the defined area of the OLED due to the high hole conductivity of the PEDOT:PSS thin film. This provides the ODMR spectrums across the entire region. (e) The most left figure (n=3) shows a zoom-in view of a local region (10 × 10 super-pixels) of the 2D map in (d). The other four figures show the magnetic field map of the same region with different binning sizes.

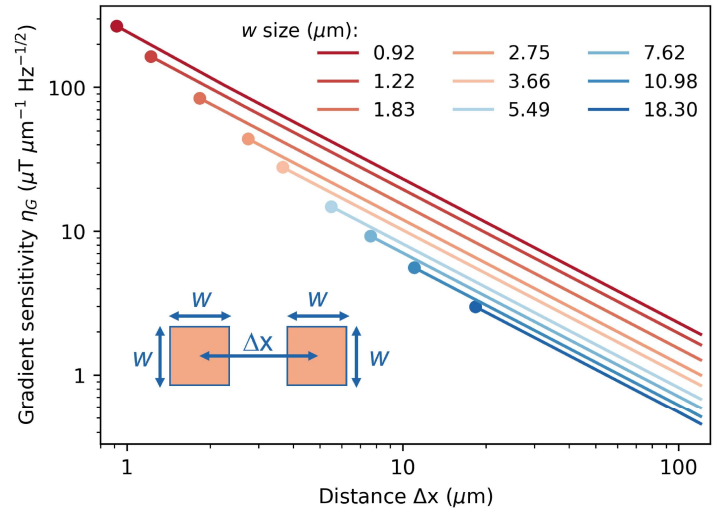

**Figure 4: Magnetic field gradient sensitivity.** The inset shows two virtual point sensors with given size of $w \times w$ and gap distance of $\Delta x$ ($\Delta x \geq w$). The dots at the left end of each curve represent the starting position where $\Delta x = w$, indicating the spatial resolution limit of the field gradient. The figure is in log-log scale plot.

# SUPPLEMENTARY INFORMATION

# Sub-micron spin-based magnetic field imaging with an organic light emitting diode

Rugang Geng, Adrian Mena, William J. Pappas, Dane R. McCamey *

ARC Centre of Excellence in Exciton Science, School of Physics, UNSW Sydney, NSW 2052, Australia

*dane.mccamey@unsw.edu.au

**Supplementary Method 1: Fabrication of the integrated microwave resonator**

Device structure and layer stack developed for this study are based on the requirement to integrate a micron-size OLED with a microwave resonator on the same substrate that provides capability for both EDMR and ODMR measurements at room temperature, good electrical insulation between the OLED from the resonator, and an open ITO surface for the implementation of the micron-size OLED on top of it. To extract the light out of the device from the bottom ITO electrode side, the resonator layer needs to be laterally separated from the ITO electrode. Gold (Au) is an excellent room-temperature conductor and resistant to most acids/chemicals, so we use Au for the resonator layer. The key procedure of integrating the resonator with an OLED on the same ITO-based glass substrate is how to electrically isolate the resonator layer from the two electrode layers of the OLED, which are the bottom ITO electrode as anode, and the top Al electrode as cathode. Here we employ low-temperature atomic-layer-deposition (ALD) method for the insulating layer deposition[1]. Unlike thermal evaporation and e-beam evaporation which have poor step-edge coverage, ALD method is conformal and provides very high-quality insulating layer even with small layer thickness, and this is essential to electrically isolate some areas with sharp edges.

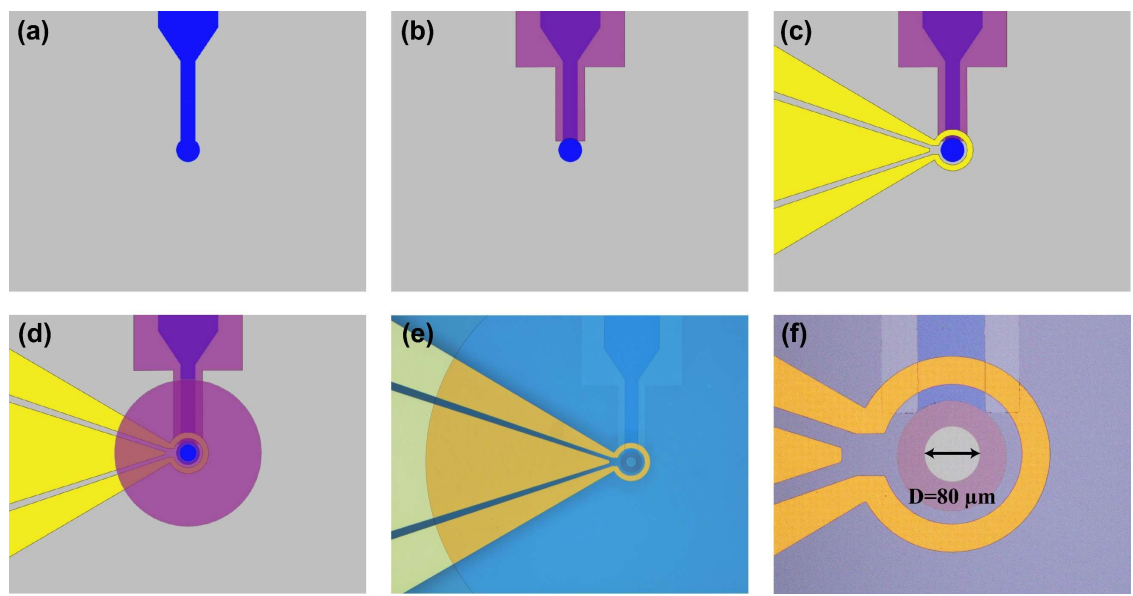

**Supplementary Figure 1:** **(a)** Patterned ITO on the glass substrate. The diameter of the electrode end-circle is about 120 μm. **(b)** First insulating layer on top of ITO electrode. **(c)** Microwave omega-shape resonator. The end of ITO electrode sits in the middle of the omega-shape centre, and the intersection between the resonator and the ITO is electrically isolated by the first insulating layer. **(d)** Second insulating layer on top of the Au resonator. The opening hole in the middle of the insulating layer defines the geometry of the active area of the OLED. **(e)** Photograph of the resonator-integrated substrate under 20× magnitude. The right side of the resonator shows more yellowish colour, indicating the area covered by the second insulating layer. Please note that the actual size of the second insulating layer is not scaled with the sketch in (d). **(f)** Zoom-in view of the resonator under 50× magnitude, and it shows a clear active area with diameter of 80 μm in the centre of the resonator. The inner diameter of the omega-shape resonator is ~ 200 μm.

The whole layer structure is as following: glass (0.7 mm) / ITO (120 nm) / $Al_2O_3$ (45 nm) / Ti (10 nm) / Au (500 nm) / Ti (10nm) / $Al_2O_3$ (45 nm).

The prepatterned ITO/glass substrates are purchased from Kintec Company (Hong Kong). The ITO substrates are cleaned by standard cleaning procedure and dried out in the vacuum drying oven at 120 °C overnight before usage. The glass substrate dimension is: 30.0 (±0.05) mm × 20.0 (±0.05) mm × 0.7 (±0.01) mm. We note that shadow mask, which will be used for thermal deposition of the top Al electrode of OLED, is precisely cut by laser based on the substrate dimension. The precision of the shadow mask dimension is particularly important for the good alignment between the substrate and the shadow mask, which is the key step in the later OLED fabrication process (see Supplementary Fig. 2).

The photoresist structures for the two insulating layers and the microwave resonator layer are prepared through the standard photolithography process using MA6 system, using negative photoresist nLOF2020 and developer ZA826MIF with optimized parameters. The details of the photolithography steps are as following:

1) Spin nLOF2020 on the substrate at 3000 RPM for 30 s, resulting in photoresist layer thickness of ~2.3 μm
2) Prebake the photoresist at 115 °C for 1 min
3) UV exposure for 4.5 s
4) Post exposure bake (PEB) the photoresist at 115 °C for 1 min
5) Develop in AZ826MIF for 1 min
6) DI water rinse for 20 s, and nitrogen gun dry
7) Further bake at 115 °C for 2 mins to remove any water residue
8) Post plasma cleaning for 10 mins (plasma etching rate ~ 30 nm / per min)

Here we choose $Al_2O_3$ as the insulating material because of its excellent electrical isolation property, more importantly, its compatibility with the materials and fabrication methods that are used in this work. The breakdown field of $Al_2O_3$ by ALD at room temperature is about 8 MV/cm (or 0.8V/nm), so 45 nm thickness is thick enough for OLEDs, whose operational voltage is in range of 0 V- 15 V. The ALD system is CNT Savannah S200. The precursors for $Al_2O_3$ in ALD are water vapor ($H_2O$) and Trimethylaluminum (TMA). The chamber temperature for the ALD process is set at 120 °C. We note that the chamber temperature cannot be set too high as it would solidate the photoresist and the following lift-off process will become exceedingly difficult. The temperature in principle can be lower such as 80 °C, which will ease the following lift-off process, but the cycle time will increase, and the total deposition time will increase dramatically. There is a trade-off between the deposition temperature and deposition time cost. The total deposition time for 45 nm $Al_2O_3$ by ALD at 120 °C is about 9.5 hours.

Following up the ALD, the lift-off procedure of $Al_2O_3$ is carried out by immersing the samples in the N-Methyl-2-pyrrolidone (NMP) bath. To allow the NMP to penetrate the conformal insulating layer and attack the photoresist below quicker, it is necessary to scratch the surface of sample manually and slightly at locations without pattern features. For the bottom insulating layer, we could easily scratch the surface close to the edge of the substrate as there is no patterns underneath; while for the top insulating layer, we employ a Cascade probe station and use the sharp metal probe-tip to crack the photoresist pillar inside the resonator gently top downwards. After the scratch, samples are immerged in the NMP bath on hotplate at 100 °C in a fume cupboard, until the lift-off procedure is completed.

For the resonator layer deposition, the substrates with prepatterned photoresist structure are transferred to a thermal evaporation chamber (Jurt J. Lesker) for the metal deposition. The vacuum condition is of $~10^{-6}$ mbar, and the layers stack is Ti (10 nm) / Au (500 nm) / Ti (10nm). The first 10 nm Ti layer is deposited as adhesion layer for the following Au deposition onto the glass surface. For the Au layer deposition, the first 100 nm is deposited with a low rate of 0.5 A/s, to minimize the heating effects on the prepatterned photoresist structure, such as deforming or softening; the next 400 nm is deposited with a high rate of 2 A/s for time saving. The second 10 nm Ti layer is deposited as another adhesion layer for the spin-coating photoresist in the following photolithography procedure. Standard lift-off is followed in the NMP bath at 100 °C.

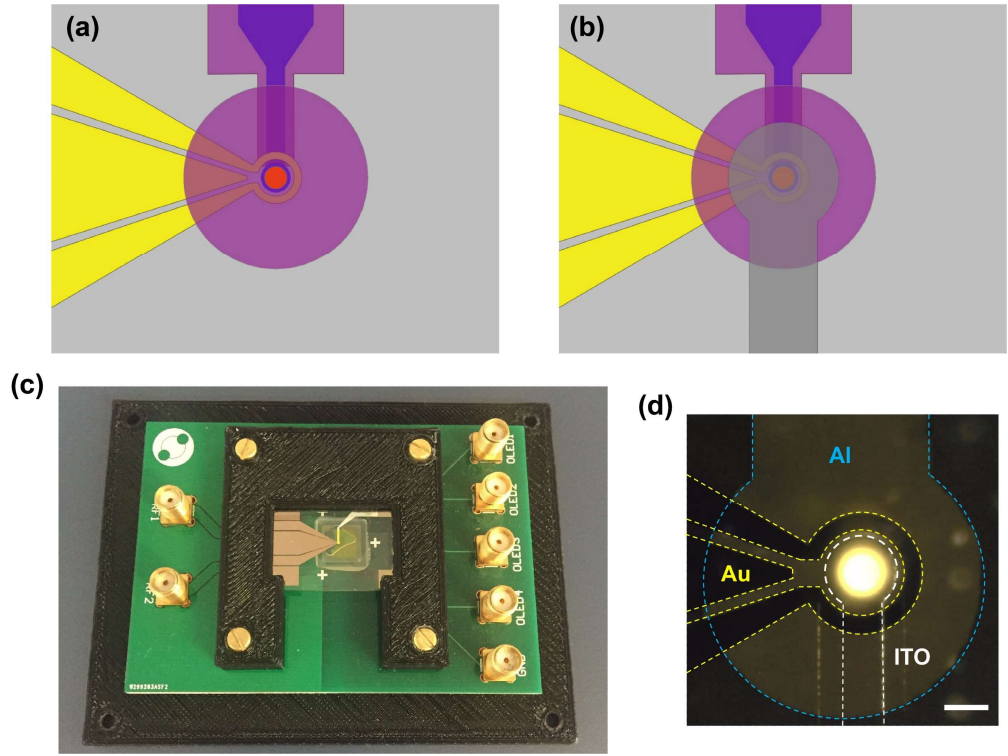

**Supplementary Figure 2:** sketch of the device fabrication **(a)** where a micron-size OLED is fabricated inside the active area (D ~ 80 μm) and **(b)** the top Al electrode is deposited by using a well aligned shadow mask. **(c)** Photograph of the PCB platform for the device mounting and electrical connection. The device is mounted onto the PCB via a 3D printed plastic lid. The device is electrically connected to the PCB via pogo pins (both AC for the resonator and DC for the OLED). The OLED is encapsulated by using a square glass coverslip (10 mm × 10 mm) with a cavity (300 μm depth) in it to avoid physical contact with the top Al electrode. **(d)** Photograph of the device under operation. Profiles of the bottom ITO electrode, the top Al electrode, and the Au resonator are highlighted using dashed lines. There is a small offset of the top Al electrode from the centre, which is due to the manual alignment of the shadow mask through the OLED fabrication procedure. The scale bar in (d) is 100 μm.

**Supplementary Method 2: Background noise caused by the microwave frequency sweep**

There is a "wobbling" feature noise in the raw magnetic resonant spectrum with sweeping the microwave frequency. We suspect that the noise origins from the microwave component in the experimental setup as we see similar "wobbling" feature in the $S_{11}$ curve of the SMA cable, which might be due to the mismatch of 50 Ω impedance of the RF output, resulting in some end-reflection of the signal in the cable. Such noise is then transmitted and modified through the PCB and the resonator, and eventually coupled to the OLED, resulting in those "wobbling" feature noise in the final EDMR spectrum (Supplementary Fig. 4). In addition, the amplitude of the noise is not a constant, but varies at different microwave frequency (Supplementary Fig. 3). Such baseline noise can be measured separately and then subtracted from the raw experimental data (Supplementary Fig. 4).

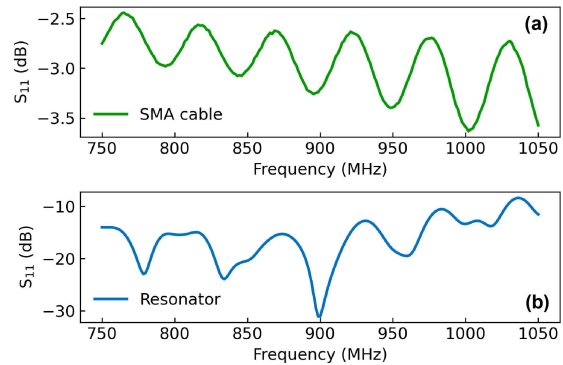

**Supplementary Figure 3:** $S_{11}$ curve of **(a)** SMA cable itself, which is connected to the microwave source directly, and **(b)** the microwave resonator connected to the microwave source through the PCB, showing a resonant frequency around 900 MHz.

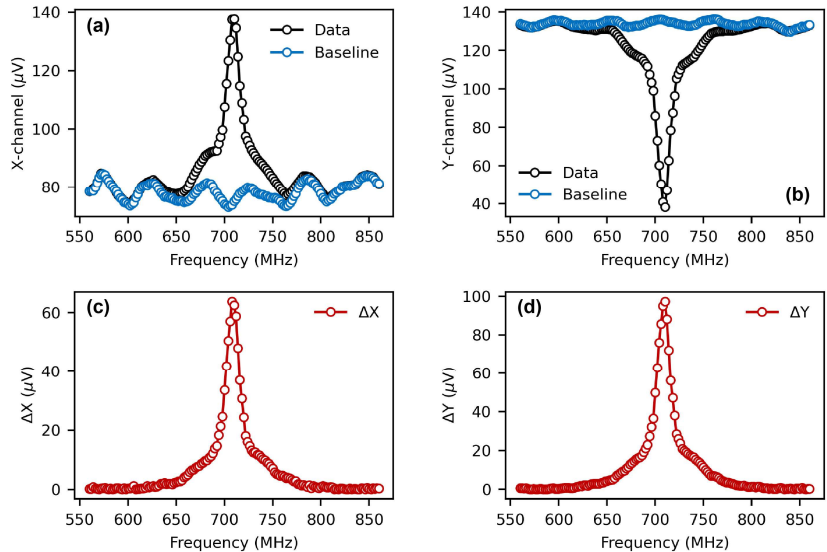

**Supplementary Figure 4:** Raw signal and background noise in **(a)** x-channel and **(b)** y-channel in the lock-in detection where the microwave frequency is swept. The signal in (a) X-channel and (b) Y-channel after the noise is subtracted. The final EDMR signal is given by $\Delta R = \sqrt{(\Delta X)^2 + (\Delta Y)^2}$ with negative sign[2,3].

**Supplementary Method 3: Angle dependent EDMR**

To verify that the magnetic resonance condition is independent of the orientation of the external magnetic field $B_0$ in our device, a separate angle dependent EDMR measurement is carried out. As sketched in Supplementary Fig. 5(a), the device is mounted on a rotation stage sitting between two electromagnet poles. Instead of changing the orientation of the field $B_0$, the orientation of microwave field $B_1$ is changed equivalently. When the device rotates, the orientation of the microwave field $B_1$ rotates in the horizontal x-y plane, labelled by angle θ as shown in Supplementary Fig. 5(b). We find that: (1) the amplitude of the EDMR spectrum peak, which corresponds to the maximum change of the EDMR signal, is proportional to the projection of the $B_1$ field along the orthogonal direction of $B_0$, and the relationship between the amplitude and the rotation angle can be well fit by a sine wave function; (2) the resonant frequency of the EDMR spectrum peak, which corresponds to the external magnetic field $B_0$ ($f=B_0\times\gamma$), remains the same with 99.94% confidence.

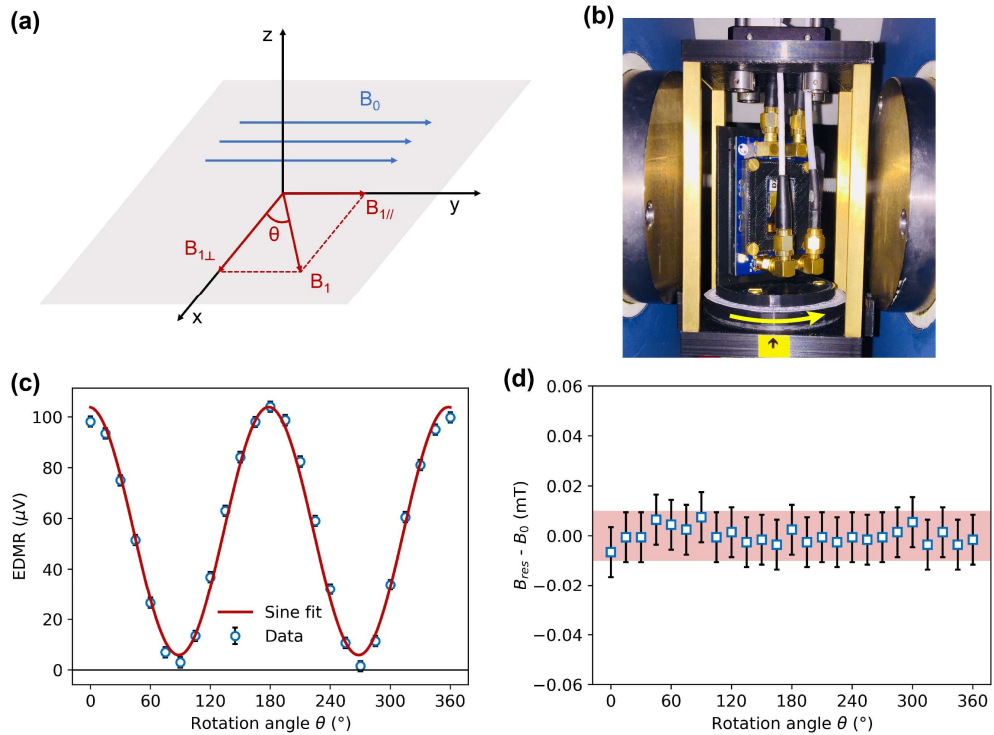

**Supplementary Figure 5: (a)** Rotation scheme where $B_0$ is the static magnetic field generated by the two electromagnet poles with fixed orientation, and $B_1$ is the oscillating field generated by the integrated microwave resonator on the device. The orientation of field $B_1$ rotates in the horizontal x-y plane along with the stage rotating. **(b)** Photograph of the experimental setup. Device is mounted on PCB vertically, and the PCB is mounted on a rotation stage. **(c)** Amplitude of the resonant peak of the EDMR spectrum as a function of the rotation angle θ, where the frequency of the microwave field $B_1$ is fixed at 700 MHz and the magnetic field $B_0$ is swept. **(d)** The difference of resonant $B_0$ field between the experimental value extracted from the EDMR spectrum and the theoretical value calculated by $B_0 = f/\gamma$, where $f$ = 700 MHz and $\gamma$ = 28.03 (GHz/T). The error bar corresponds to the magnetic field step accuracy (0.02 mT) in this measurement.

We also observed an exceedingly small EDMR signal at $\theta = 90°$ and $\theta = 270°$, which is attributed to the spatial variation of the microwave field. The distance between the microwave resonator and the OLED is much smaller than the wavelength of the microwave radiation, hence the OLED is in the near-field region of the $B_1$ field. As a result, the spatial orientation of the $B_1$ field is determined by the dimensions of the resonator itself and the surrounding conductors, and the orientation varies slightly in the OLED region. Therefore, there is always a small in-plane projection ($B_{1//}$) of the $B_1$ field during the whole rotation, and it plays a dominant role in the non-zero EMDR spectrums at $\theta = 90°$ and $\theta = 270°$.

The small variation of resonant $B_0$ field value origins from the following aspects: (1) The influence of the SMA connectors on the PCB. We find that those SMA connectors used in this work show weak paramagnetic behaviour under large external magnetic field, and it leads to a very weak disturbance on the Gauss probe reading during the rotation. The disturbance becomes noticeable (at 0.1 mT scale) at some angles (60° to 120°) where the SMA connectors are closest to the Gauss probe. We note that such disturbance has been removed in Supplementary Fig. 5(d) through an independent and careful calibration process. (2) The finite step size of the sweeping magnetic field, which results in an uncertainty of the $B_0$ value extraction. (3) The uniformity of the static magnetic field $B_0$ between the two electromagnet poles. The uniformity of the field depends on the dimensions of the two poles, the gap between them, and the spatial location. In practice, the magnetic field detected by the Gauss probe is always slightly different from the actual field the device experiences, and such difference may even vary during the rotation as the rotation setup is not perfectly aligned with the magnet.

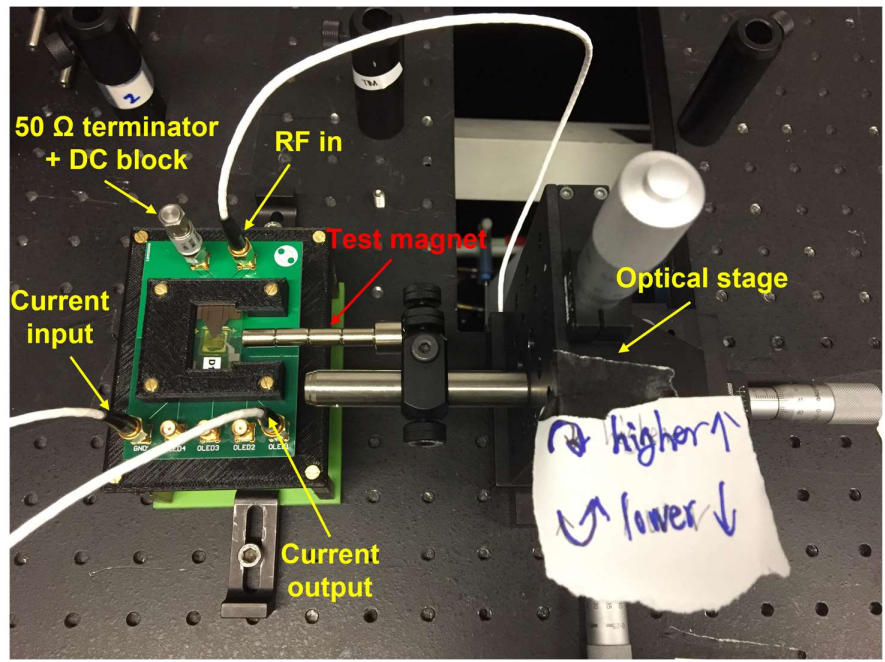

**Supplementary Figure 6:** Setup for frequency-swept-based EDMR measurement.

# Supplementary Method 4: Simulation of the magnetic field

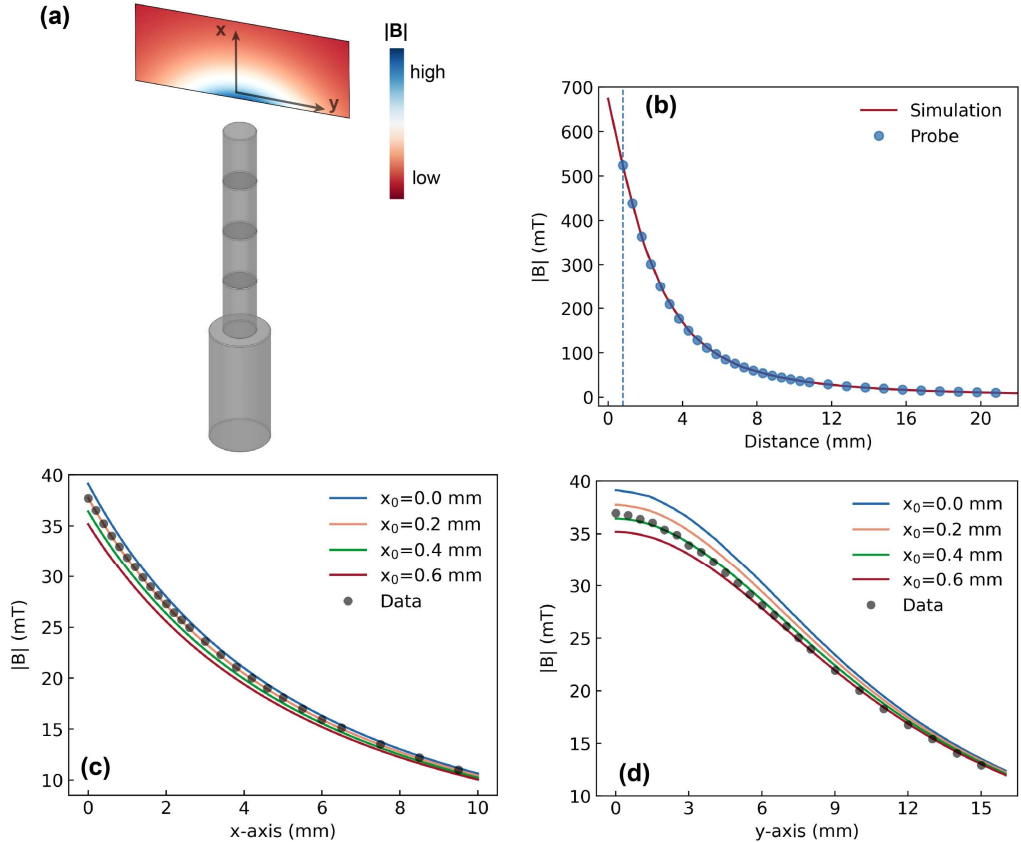

**Supplementary Figure 7:** **(a)** Magnetic field simulation of the test magnets using ANSYS Electronics (Maxwell 3D Design with Magnetostatic Solution). The 2D map represents the spatial distribution of the strength of the static magnetic field generated by the magnet in a region of 14.0 mm × 36.0 mm in the x-y plane with a gap distance of d = 10.0 mm from the top surface of the magnet. Please note that the test magnets are not drawn to scale. Dimensions and material properties of these cylindrical magnets are summarized in Supplementary Table 1. **(b)** Comparison of the simulated magnetic field and the experimentally measured field using Gauss probe. The strength of the magnetic field here is a function of the movement distance along the z-direction from the top surface of the magnet. We note that the total thickness of the Hall probe is about 1.6 mm, therefore the starting position for the Hall probe measurement is about 0.8 mm (half of the probe thickness) as labelled out by the dash line, where we positioned the probe right adjacent to the magnet. The similarity between the Hall probed field and the simulation field is 98.5 %. Comparison of experimentally measured field (grey dots) and the simulated magnetic field (colour curves) with various starting position $x_0$, as a function of the movement distance along x-direction in **(c)** and along y-direction in **(d)**, respectively. We note that the x-y coordinates in (c) and (d) is a local frame within the 2D map plane, which is labelled out separately in (a). Similarity between the measured field and the simulation field is 99.8 % ($x_0$ = 0.2 mm) in (c), and 98.6 % ($x_0$ = 0.4 mm) in (d). The similarity calculation formula is listed as below.

| | Diameter (mm) | Length (mm) | Edge radius (mm) | Materials |
|---|---|---|---|---|
| **Cylinder 1 (×4)** | 7.0 | 12.0 | 0.2 | N48 |
| **Cylinder 2** | 12.7 | 25.4 | 0.2 | N45 |

**Supplementary Table 1**: We note that the parameters of diameter, length and material property are obtained directly from the product datasheet, and the edge radius is estimated based on our own measurements. The edge radius here refers to the smooth curvature of the surface edge of the magnet cylinder. In addition, the influence of the edge radius on the simulation field distribution is investigated by comparing simulation with edge radius 0.2 mm and simulation with edge radius 0.0 mm. We find: (1) for the far-field region (d>8.0 mm), the difference of the field between two cases, both amplitude and direction, is negligibly small; (2) for the near-field region (d<5.0 mm around the edge area), the difference is still quite small but not negligible. Therefore, in the Main Fig. 2 where the distance d >10 mm, simulation field remains the same regardless the estimated value of the edge radius.

- Definition of similarity: $s = 1 - \frac{1}{N} \sum_{i=1}^{N} \frac{|p(x_i) - q(x_i)|}{[p(x_i) + q(x_i)]/2}$, where $p(x_i)$ and $q(x_i)$ are two independent data sets as a function of variable $x_i$.

## Supplementary Method 5: Standard error calculation

- **Standard error in the local region**

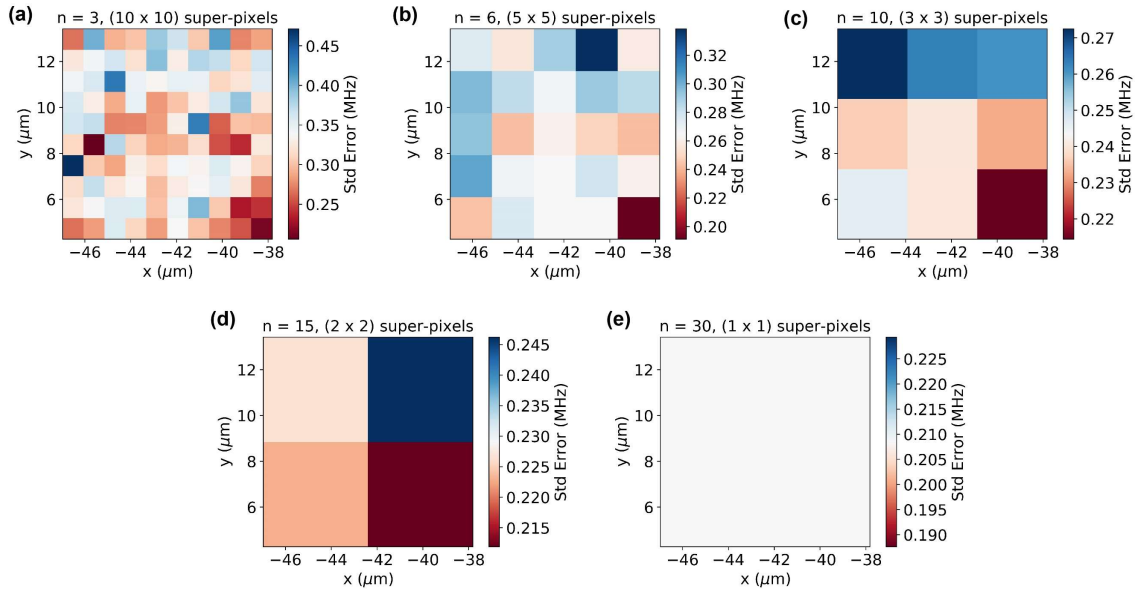

**Supplementary Figure 8:** standard error of the fit of the resonant peak frequency with a variety of binning size in the local region.

- **Standard error in the OLED region and diffusion region**

Figure 9(a) shows the spatial distribution of the standard error (SE) of the resonant peak frequency of the ODMR spectrums. The 2D map demonstrates three distinguishable regions: (1) the central region with $R < R_2$, where the SE is clearly larger than the surrounding regions. The reason of the large SE in this region is because of the electrical coupling between the resonator and the device electrodes. We suspect that the 'wobbling' feature noise from the resonator (see Supplementary Fig. 3) is encoded into the device where an electrical coupling between the ITO and Al electrodes is induced. Such electrical coupling is eventually transmitted to the device output (both current and the EL), reducing the overall SNR. (2) the ring region with $R_2 < R < R_3$, where the SE is the smallest. The EL emission in this ring region is due to the high hole mobility in the PEDOT:PSS layer. In specific, holes are injected from ITO electrode into PEDOT:PSS layer through the defined area ($R_{OLED} = 40$ μm), then diffuse in the PEDOT:PSS layer along the in-plane direction outwards. Under the bias voltage, these diffusing holes are gradually injected into the emitting layer along the diffusion path, eventually forming excitons through combing with the electrons which are injected from the top Al electrode. As this is the diffusion region, the noise caused by the electrical coupling between the two electrodes is much weaker compared to the central region. Therefore, the SE in this ring region is smaller than the central region. (3) the edge region with $R > R_3$, where the SE is large compared to ring region though it is also the diffusion region. The reason is because the EL signal is much weaker in this edge region, therefore the over SNR is much smaller.

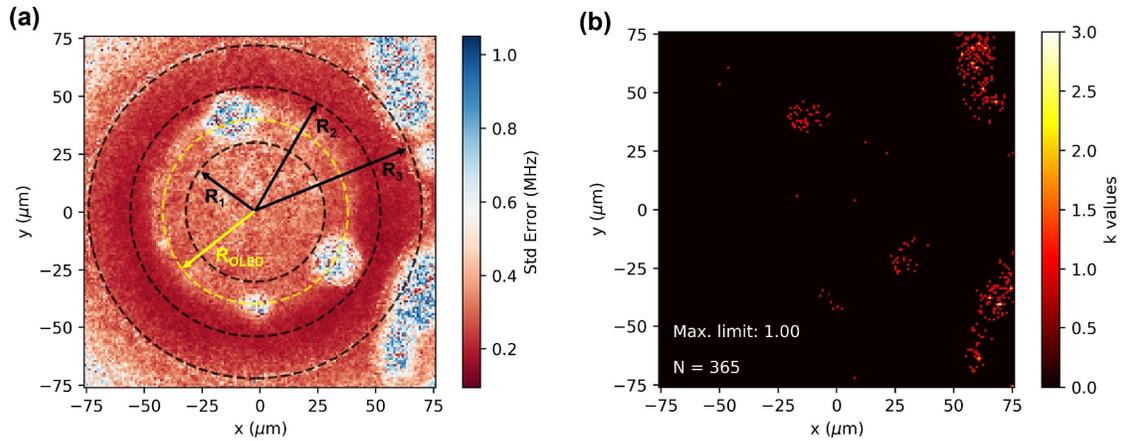

**Supplementary Figure 9:** **(a)** 2D map of the standard error of the resonant peak frequency fit with binning size n =3 (the whole region contains of 166 × 166 super-pixels). The OLED region refers to the region with R < $R_1$ and the diffusion region refers to the region with $R_2$< R < $R_3$. The yellow dash circle with radius $R_{OLED}$ indicates the edge of the OLED area defined by the photolithography process. Their values are: $R_1$= 30 µm, $R_2$ =54 µm, $R_3$ = 72 µm, $R_{OLED}$ = 40 µm. **(b)** Spatial distribution of outliers which are defined as the points whose value is out of the field range of (754 MHz, 771 MHz). The actual values of the resonant peak frequency and the standard error of those outliers are interpolated by adjacent points, the distance of which is indicated by k value. The selection of the values of those radiuses is to avoid the outliers while covering as many points as possible in each region. Those outliers are caused by defects during photolithography process and device degradation during the measurement.

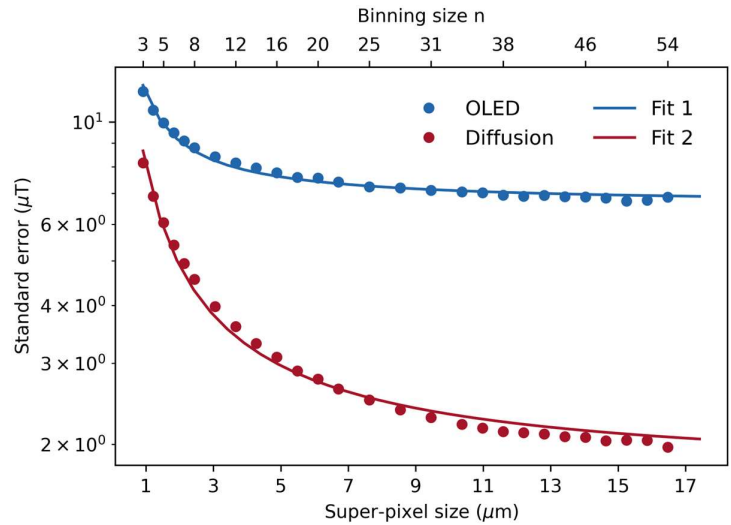

**Supplementary Figure 10:** Standard error (SE) of double Gaussian fits of the spatially resolved ODMR spectrums with different binning sizes.

**Supplementary Method 6: Sensitivity calculation of the spatially resolved ODMR**

- **Magnetic field sensitivity**

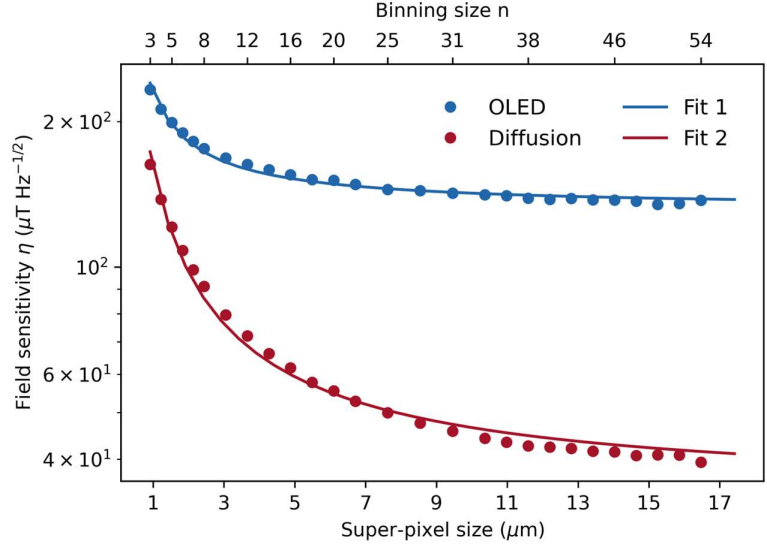

**Supplementary Figure 11:** Magnetic field sensitivity of the spatially resolved ODMR as a function of super-pixel size in both OLED and diffusion regions. See more details about the region definition in Supplementary Figure 9. The data set can be fit by using function $y = a/x + b$. The fit parameters are (1) for the OLED region: $a = 98.81(\pm2.12)$, $b = 132.33(\pm0.76)$; (2) for the diffusion region: $a = 127.55(\pm2.44)$, $b = 33.77(\pm0.87)$. The unit for parameter $a$ and $b$ is $\mu T \cdot Hz^{-1/2} \cdot \mu m$ and $\mu T \cdot Hz^{-1/2}$, respectively.

- **Magnetic field gradient sensitivity**

The magnetic field gradient sensitivity is calculated from the uncertainty of the field gradient, which is given by $G = \Delta B/\Delta x$, where $\Delta B = B(x_2) - B(x_1)$ is the measured magnetic field difference between two super-pixels located at $x_1$ and $x_2$, and $\Delta x = x_2 - x_1$ is the distance between them (refers to the inset in Main Figure 4). Based on the error propagation, the minimum detectable gradient $\delta G$ is given as following:

$$\frac{(\delta G)^2}{G^2} = \frac{(\delta \Delta B)^2}{(\Delta B)^2} + \frac{(\delta \Delta x)^2}{(\Delta x)^2} \quad (1)$$

Because of $\Delta B = B(x_2) - B(x_1)$, So $(\delta \Delta B)^2 = [\delta B(x_2)]^2 + [\delta B(x_1)]^2$, where $\delta B(x_i) (i=1,2)$ is the uncertainty of the magnetic field (or minimum detestable magnetic field). Here we assume that uncertainty of the field is location independent or ideally the same across the whole device, $\delta B(x_2) = \delta B(x_1) = \delta B$, therefore we have the following result:

$$\delta \Delta B = \sqrt{2} \times \delta B \quad (2)$$

Similarly, we can have the following result:

$$\delta\Delta x = \sqrt{2}\times\delta x \qquad (3)$$

where $\delta x$ is the uncertainty of the distance measurement and we assume $\delta x$ is location independent as well.

For a digital distance measurement, the minimum uncertainty is equal to the super-pixel width (see the inset in figure 4 in the main content),

$$\delta x = w \qquad (4)$$

As the whole measurement system is fixed firmly on the optical table, and no relative movement between the camera and device has been observed in a similar setup from our previous work[4], we believe that the actual uncertainty of the distance measurement caused by vibration and relative displacement is negligibly small compared to $w$.

By plugging equations (2-4) into equation (1) and using $G = \Delta B/\Delta x$, we can get:

$$\frac{(\delta G)^2}{G^2} = \frac{2(\delta B)^2}{(\Delta B)^2} + \frac{2(\delta x)^2}{(\Delta x)^2}$$

$$= G^2\left[\frac{2(\delta B)^2}{(\Delta B)^2} + \frac{2(\delta x)^2}{(\Delta x)^2}\right]$$

$$= G^2\frac{2(\delta B)^2}{(\Delta B)^2} + G^2\frac{2(\delta x)^2}{(\Delta x)^2}$$

$$= \frac{(\Delta B)^2}{(\Delta x)^2}\frac{2(\delta B)^2}{(\Delta B)^2} + G^2\frac{2(\delta x)^2}{(\Delta x)^2}$$

$$= \frac{2(\delta B)^2}{(\Delta x)^2} + \frac{2G^2(\delta x)^2}{(\Delta x)^2}$$

$$= \sqrt{2}\frac{\delta B}{\Delta x}[1+(\frac{\delta x}{\delta B})^2 G^2]^{1/2} \qquad (5)$$

Equation (5) can be rewritten as below:

$$(\delta G) = \sqrt{2}\frac{\delta B}{\Delta x}[1+(\frac{\delta x}{\delta B})^2(\frac{\Delta B}{\Delta x})^2]^{1/2} \qquad (6-1)$$

$$(\delta G) = \sqrt{2}\frac{\delta B(w)}{\Delta x}\{1+[\frac{w}{\delta B(w)}]^2(\frac{\Delta B}{\Delta x})^2\}^{1/2} \qquad (6-2)$$

In the above equation (6-1), $\delta B$ is the minimum detectible magnetic field (or uncertainty of the magnetic field as mentioned above), and it depends on the super-pixel size $w$ (or binning size $n$) as shown in Supplementary Figure 10; $\delta x = w$ is the uncertainty of the distance measurement as shown in Equation (4). Equation (6-1) can also be rewritten as equation (6-2) which shows all the independent variables: $w$, $\Delta x$, and $\Delta B$.

According to equation (6-2), $\delta G$ is a monotonic increasing function of variable $\Delta B$ with fixed value of $w$ and $\Delta x$. Therefore, the minimum detectable gradient $\delta G$ can be further decreased by choosing the minimum field difference $(\Delta B)_{min} = \sqrt{2} \times \delta B$ as shown in equation (2). By plugging $(\Delta B)_{min} = \sqrt{2} \times \delta B$ into equation (6-2), we can get $\delta G$ as below:

$$(\delta G) = \sqrt{2}\frac{\delta B(w)}{\Delta x}[1+2(\frac{w}{\Delta x})^2]^{1/2} \quad (\Delta x \geq w) \tag{7}$$

Finally, the magnetic field gradient sensitivity $\eta_G$ is given as below:

$$\eta_G = (\delta G)\sqrt{T}$$

$$= \sqrt{2T}\frac{\delta B(w)}{\Delta x}[1+2(\frac{w}{\Delta x})^2]^{1/2} \tag{8}$$

Please note that in the equation (8) all the uncertainties (or errors) should be relatively small to justify the error propagation analysis above.